\documentclass[aps,prc,superscriptaddress,showpacs,floatfix,nofootinbib,notitlepage,twocolumn]{revtex4-1}
\usepackage{amsmath,graphicx,float,hyperref,upgreek}

\begin{document}

\title{Thermalization at the femtoscale seen in high-energy Pb+Pb collisions}
\author{Rupam Samanta}
\affiliation{AGH University of Science and Technology, Faculty of Physics and
Applied Computer Science, aleja Mickiewicza 30, 30-059 Cracow, Poland}
\affiliation{Universit\'e Paris Saclay, CNRS, CEA, Institut de physique th\'eorique, 91191 Gif-sur-Yvette, France}
\author{Somadutta Bhatta}
\affiliation{Department of Chemistry, Stony Brook University, Stony Brook, NY 11794, USA}
\author{Jiangyong Jia}
\affiliation{Department of Chemistry, Stony Brook University, Stony Brook, NY 11794, USA}
\affiliation{Physics Department, Brookhaven National Laboratory, Upton, NY 11976, USA}
\author{Matthew Luzum}
\affiliation{Instituto de F\'{\i}sica, Universidade de  S\~{a}o Paulo,  Rua  do  Mat\~{a}o, 1371,  Butant\~{a},  05508-090,  S\~{a}o  Paulo,  Brazil}
\author{Jean-Yves Ollitrault}
\affiliation{Universit\'e Paris Saclay, CNRS, CEA, Institut de physique th\'eorique, 91191 Gif-sur-Yvette, France}

\begin{abstract}
A collision between two atomic nuclei accelerated at a speed close to that of light creates a dense system of quarks and gluons. 
Interactions among them are so strong that they behave collectively like a droplet of fluid of ten-femtometer size, which expands into the vacuum and eventually fragments into thousands of particles. 
We report a new manifestation of thermalization in recent data from the Large Hadron Collider. 
Our analysis is based on results from the ATLAS Collaboration, which has measured the variance of the momentum per particle across Pb+Pb collision events with the same particle multiplicity. 
This variance decreases steeply over a narrow multiplicity range corresponding to central collisions. 
We provide a simple explanation of this newly-observed phenomenon: 
For a given multiplicity, the momentum per particle increases with increasing impact parameter. 
Since a larger impact parameter goes along with a smaller collision volume, this in turn implies that the momentum per particle increases as a function of density, which is a generic consequence of thermalization.  
%This means that there are interactions among the created particles, which lead to thermalization. 
Our analysis provides the first direct evidence of this phenomenon at the femtoscale.  
\end{abstract}

\maketitle
Nucleus-nucleus collisions carried out at particle colliders display phenomena of macroscopic nature, which are unique in the realm of high-energy physics~\cite{Busza:2018rrf,Schenke:2021mxx}.  
These emergent phenomena occur due to a large number of created particles and to the nature of the strong interaction. 
A head-on collision between two $^{208}$Pb nuclei at the Large Hadron Collider (LHC), at 5.02~TeV per nucleon pair (the current energy for ion beams), produces some 35000 hadrons~\cite{ALICE:2016fbt}, a fraction of which are seen in detectors. 
The emission of hadrons is the final outcome of a number of successive stages~\cite{Busza:2018rrf}, one of which is the production of a state of matter called the quark-gluon plasma. 
In this phase, quarks and gluons, which are the elementary components of hadrons, are liberated~\cite{Gardim:2019xjs}. 
They carry color charges, unlike hadrons which are colorless. 
Interactions induced by these charges are so strong that they behave collectively like a fluid~\cite{Shuryak:2003xe}. 

\begin{figure}[h]
\begin{center}
\includegraphics[width=0.7\linewidth]{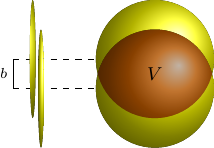} 
\end{center}
\caption{
  Schematic representation of a collision between two identical spherical nuclei at impact parameter $b$.
  Left: Incoming nuclei just before the collision, flattened by the relativistic Lorentz contraction in the direction of motion.
  Right: View from the collision axis. 
Strongly-interacting matter is created in the region where the nuclei overlap, which is indicated in darker color, and $V$ is the collision volume.
Generally, smaller $b$ is associated with larger $V$.
}
\label{fig:cartoon}
\end{figure}

Transient formation of a fluid in nucleus-nucleus collisions has been inferred from the observation that particles move collectively into preferred directions, suggesting that their motion is driven by pressure gradients inherent in a fluid. 
Most notably, one observes an elliptic deformation of the azimuthal distribution of outgoing particles~\cite{STAR:2000ekf,ALICE:2010suc},  which originates from the almond shape of the overlap area between the colliding nuclei (Fig.~\ref{fig:cartoon}). 
These observations are reproduced by calculations using relativistic hydrodynamics to model the expansion of the fluid~\cite{Gale:2013da}, which has become the standard description of nucleus-nucleus collisions. 

\begin{figure*}[h]
\begin{center}
\includegraphics[width=\linewidth]{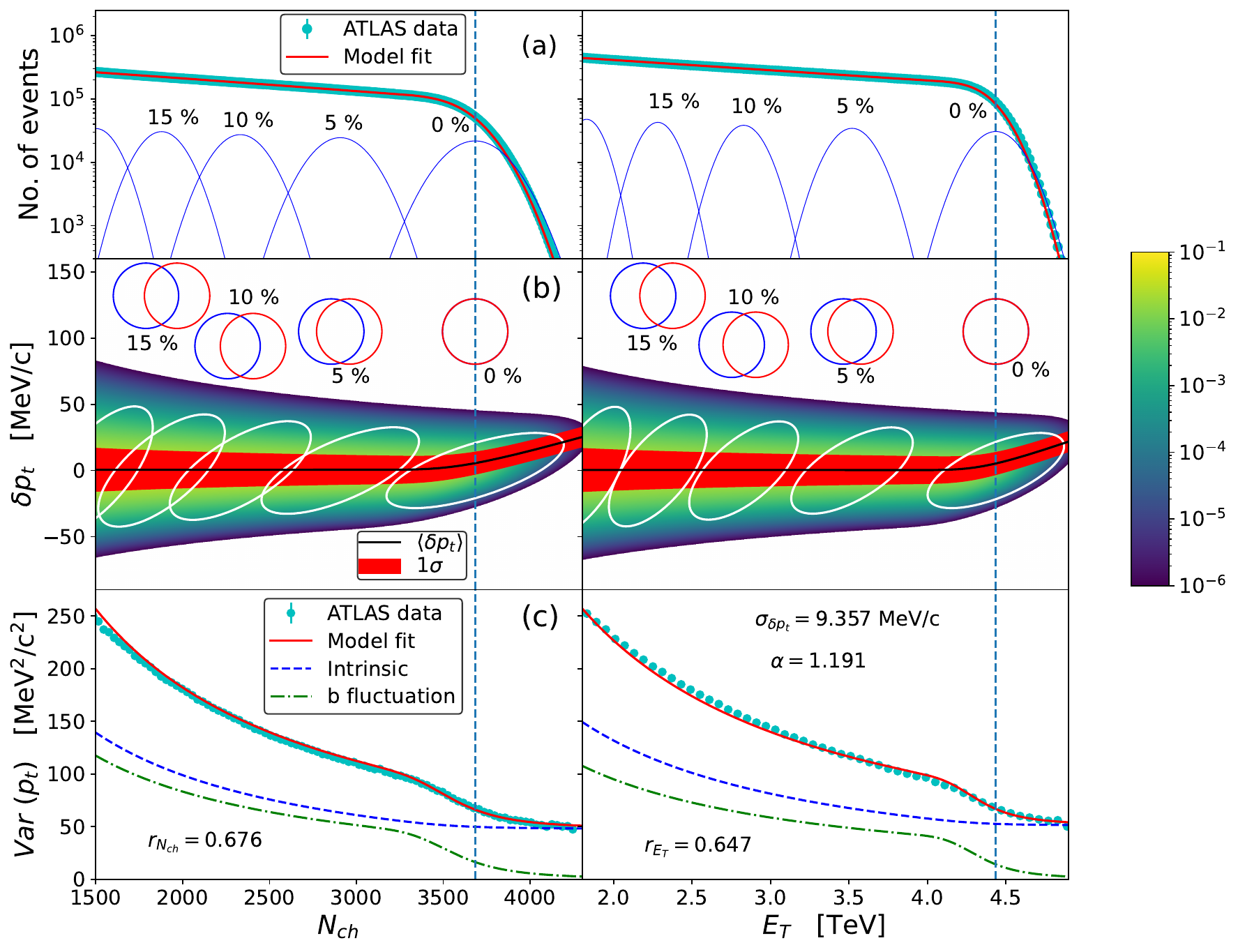} 
\end{center}
\caption{
(a)  Histogram of the number of charged particles $N_{ch}$ (left), measured in the inner detector of ATLAS, and of transverse energy $E_T$ (right), measured in the forward and backward calorimeters. 
Solid lines are fits using superpositions of Gaussians. 
Contributions of collisions at fixed impact parameter $b$ corresponding to centrality fractions $0$, $5\%$, $10\%$, $15\%$ are shown as thin blue lines (see Appendix~\ref{s:bayesian} for details).  
The vertical dashed line corresponds to the knee, defined as the average value of  $N_{ch}$ or $E_T$ for $b=0$ collisions. 
(b) Joint distribution of the transverse momentum per particle $[p_t]$ and $N_{ch}$ (or $E_T$)  from our model.
Rather than $[p_t]$, we plot the deviation $\delta p_t\equiv [p_t]-\overline{p_{t0}}$, where $\overline{p_{t0}}$ is the average value of $p_t$ at fixed impact parameter, which is assumed to be constant.  
White curves are 99\% confidence ellipses at fixed $b$. 
A schematic representation of the two colliding nuclei for these values of $b$ is also shown (Appendix~\ref{s:bayesian}). 
The black line is the mean value of $\delta p_t$, and the red band is the 1-$\sigma$ band. 
(c) Variance of the transverse momentum per particle  $[p_t]$ as a function of the centrality estimator. 
The red solid line is the square of the half-width of the red band in panel (b). 
Symbols are ATLAS data~\cite{ATLAS:2022dov}.
We also display separately the two contributions to the variance, Eq.~(\ref{decomposition}), in our model calculation, whose sum is the full line. 
}
\label{fig:panelplot}
\end{figure*} 

Here, we report independent confirmation of the formation of a fluid, which does not involve the directions of outgoing particles, but solely their momenta.  
The ATLAS Collaboration at the LHC detects charged particles in an inner detector which covers roughly the angular range $10^\circ<\theta< 170^\circ$ (where $\theta$ is the angle between the collision axis and the direction of the particle) and measures their transverse momenta $p_t\equiv p\sin\theta$. 
The analysis includes all charged particles detected in the interval $0.5<p_t<5$~GeV$/c$. 
The observables of interest are, for every collision, the multiplicity of charged particles seen in the inner detector, denoted by $N_{ch}$, and the transverse momentum per charged particle, $(\sum p_t)/N_{ch}$, denoted by $[p_t]$.
$N_{ch}$ is used to estimate the centrality~\cite{Back:2000gw,Adler:2001yq,Adler:2004zn,ALICE:2013hur}, since a more central collision, with a smaller impact parameter, produces on average more particles. 

For collisions with the same $N_{ch}$, $[p_t]$ fluctuates from event to event. 
After subtracting trivial statistical fluctuations, the remaining dynamical fluctuations~\cite{Adams:2003uw} are very small, below 1\% in central Pb+Pb collisions at the LHC~\cite{Abelev:2014ckr}. 
These small dynamical fluctuations are the focus of our study. 
The left panel of Fig.~\ref{fig:panelplot} (c) displays their variance as a function of $N_{ch}$~\cite{ATLAS:2022dov}.  
The striking phenomenon is a steep decrease, by a factor $\sim 2$, over a narrow interval of $N_{ch}$ around $3700$. 
This behavior is not reproduced by models of the collision in which the Pb+Pb collision is treated as a superposition of independent nucleon-nucleon collisions, such as the HIJING model~\cite{Wang:1991hta,Gyulassy:1994ew}, where the decrease of the variance is proportional to $1/N_{ch}$~\cite{Abelev:2014ckr,Bhatta:2021qfk} for all $N_{ch}$.

\begin{figure}[h]
\begin{center}
\includegraphics[width=0.9\linewidth]{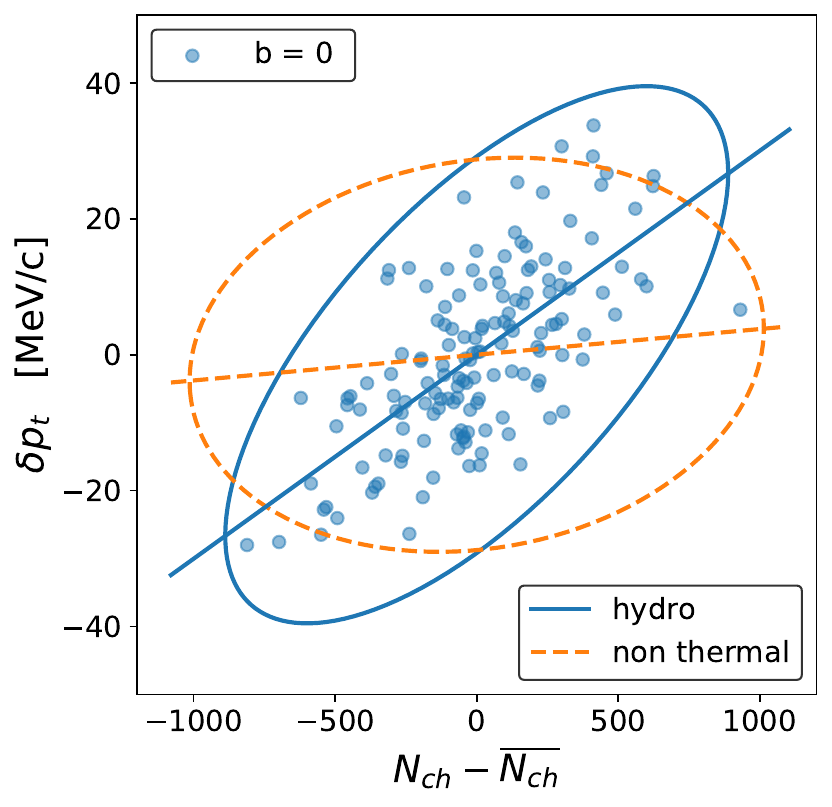} 
\end{center}
\caption{
Simulations of Pb+Pb collisions at 5.02~TeV and $b=0$. 
The first set of simulations, shown as solid lines and symbols, consists of 150 collisions modeled using relativistic hydrodynamics~\cite{Bozek:2021mov}. The second set, shown as dashed lines, consists of $1.4\times 10^6$ collisions simulated with HIJING~\cite{Gyulassy:1994ew}, in which there is no thermalization mechanism. 
The figure represents the distribution of the charged particle multiplicity $N_{ch}$ and the transverse momentum per particle $[p_t]$, where $N_{ch}$ is calculated using the same acceptance cuts on $\theta$ and $p_t$ as in the ATLAS analysis. 
We plot, rather than $N_{ch}$ and $[p_t]$ themselves, the differences $N_{ch}-\overline{N_{ch}}$ and $\delta p_t\equiv [p_t]-\overline{p_t}$,  where $\overline{N_{ch}}=6662$ and $\overline{p_t}=1074$~MeV$/c$ are the values averaged over collisions. 
The straight lines indicate the average value $\overline{\delta p_t}(N_{ch},b=0)$, and the ellipses are 99\% confidence ellipses (as in Fig.~\ref{fig:panelplot} (b)). Both are evaluated by assuming that the distribution is Gaussian (Appendix~\ref{s:gaussian}). 
Note that the fluctuations contain a contribution from statistical Poisson fluctuations in the HIJING model, which is a particle-based description, not in hydrodynamics, which is a continuous description (Appendix~\ref{s:hydro}). 
}
\label{fig:hydro}
\end{figure} 

We will argue that the impact parameter, $b$, plays a crucial role in this phenomenon. 
The relation between $N_{ch}$ and $b$ is not one-to-one, and $[p_t]$ depends on both quantities. 
In order to illustrate this dependence, we simulate 150 collisions at $b=0$ using relativistic viscous hydrodynamics, and evaluate $N_{ch}$ and $[p_t]$ for every collision. 
Figure~\ref{fig:hydro} displays their distribution. 
The first observation is that they span a finite range. 
Fluctuations around the mean extend up to $\sim 14\%$ for $N_{ch}$, and to $\sim 3\%$ for $[p_t]$. 
They originate from quantum fluctuations at different levels:  
In the positions of nucleons at the time of impact~\cite{Miller:2007ri}, in the partonic content of the nucleons~\cite{Gelis:2010nm}, and in the process of particle production.\footnote{We only consider spherical nuclei. For deformed nuclei, one must also consider fluctuations in their orientations, which affect both the multiplicity~\cite{STAR:2015mki} and the momentum per particle~\cite{Giacalone:2019pca}.} 
Modern hydrodynamic simulations take these fluctuations into account~\cite{Aguiar:2001ac} by implementing a different initial density profile  (the initial condition of  hydrodynamic equations) in every collision. 
The second observation in Fig.~\ref{fig:hydro} is that there is a positive correlation between $[p_t]$ and $N_{ch}$ in hydrodynamics. 

This correlation is a consequence of local thermalization, which is an underlying assumption of the hydrodynamic description. 
Larger $N_{ch}$ implies a larger density $N_{ch}/V$, as the volume $V$ (Fig.~\ref{fig:cartoon}) is essentially defined by the impact parameter, which is fixed. 
In hydrodynamics, one assumes that the system is locally thermalized, and larger density corresponds to higher initial temperature. 
Note that relativity plays an essential role in this correspondence. 
In non-relativistic thermodynamics, density and temperature are independent variables. 
Heating a system at constant volume does not change its density, because the number of particles is conserved. 
In a relativistic system, on the other hand, particles can be created by converting kinetic energy into mass, and a higher temperature implies a higher density. 
It also implies a higher energy per particle, which eventually results in a larger momentum per particle $[p_t]$~\cite{Gardim:2019xjs}.

In order to illustrate that the positive correlation between $[p_t]$ and $N_{ch}$ is not trivial,  Fig.~\ref{fig:hydro} also displays results of simulations using the HIJING model~\cite{Gyulassy:1994ew}, in which particles do not interact after they are produced. 
The correlation is smaller by a factor $\sim 10$. 
Note, however, that while thermalization always implies a positive correlation, the converse statement does not hold. 
In the color-glass condensate picture of high-energy collisions, such a correlation is already present at the level of particle production, since both the momentum per particle and the particle density increase with the saturation scale~\cite{Gelis:2010nm}.

We now discuss the implications of thermalization on the observed $[p_t]$ fluctuations. 
First, note that the experimental analysis is done at fixed $N_{ch}$, while our hydrodynamic simulation is done at fixed $b$. 
Both choices are dictated by practical reasons. 
Experimentally, $b$ is not measured. 
In the simulation, on the other hand, one must define $b$ before starting the simulation, while $N_{ch}$ is only evaluated at the end. 

In order to understand experimental results, we must reason at fixed $N_{ch}$, where $b$ varies. 
Larger $b$ implies smaller collision volume $V$ and larger density $N_{ch}/V$, hence larger $[p_t]$ on average. 
We denote by $\overline{p_t}(N_{ch},b)$ the expectation value of $[p_t]$ at fixed $N_{ch}$ and $b$. 
It increases with $N_{ch}$ at fixed $b$, and with $b$ at fixed $N_{ch}$. 
In addition, there are fluctuations of $[p_t]$ even if both $N_{ch}$ and $b$ are fixed, as illustrated by the simulation in Fig.~\ref{fig:hydro}. 
We denote by ${\rm Var}(p_t|N_{ch},b)$ their variance. 
We then average over $b$ at fixed $N_{ch}$. 
The average value of $[p_t]$ is $\left\langle\overline{p_t}(N_{ch},b)\right\rangle_b$, where $\langle\cdots\rangle_b$ denotes an average over $b$.
The average value of $[p_t]^2$ is $\left\langle\overline{p_t}(N_{ch},b)^2+{\rm Var}(p_t|N_{ch},b)\right\rangle_b$.
Therefore, the variance of $[p_t]$ is the sum of two positive terms:
\begin{eqnarray}
  \label{decomposition}
        {\rm Var}(p_t|N_{ch})&=&\left(\left\langle\overline{p_t}(N_{ch},b)^2\right\rangle_b-\left\langle\overline{p_t}(N_{ch},b)\right\rangle_b^2\right)\cr
        &&+\left\langle{\rm Var}(p_t|N_{ch},b)\right\rangle_b,
\end{eqnarray}
The first term stems from the variation of $\overline{p_t}(N_{ch},b)$ with $b$. 
We refer to the second term as the intrinsic variance, in the sense that it is not a by-product of $b$ fluctuations.
As we shall see, both terms are of comparable magnitudes, and the first term explains the peculiar pattern observed for large $N_{ch}$.  

We now carry out a quantitative calculation, which can be compared with data. 
First, precise information can be obtained, without any microscopic modeling, about the probability distribution of $b$ at fixed $N_{ch}$,  $P(b|N_{ch})$~\cite{Das:2017ned}. 
This is achieved by solving first the inverse problem, namely, finding the probability distribution of $N_{ch}$ for fixed $b$, $P(N_{ch}|b)$, and then applying Bayes' theorem $P(b|N_{ch})P(N_{ch})=P(N_{ch}|b)P(b)$. 
As explained above, collisions at the same $b$ differ by quantum fluctuations, which result in fluctuations of $N_{ch}$. 
In nucleus-nucleus collisions, these fluctuations are Gaussian to a good approximation. 
They are characterized by the mean, $\overline{N_{ch}}(b)$, and the variance, ${\rm Var}(N_{ch}|b)$.  

What one measures is the distribution $P(N_{ch})$, obtained after integrating over all values of $b$, shown in Fig.~\ref{fig:panelplot} (a), left.
We only display values of $N_{ch}$ larger than some threshold such that only 20\% of the events are included, corresponding to fairly central collisions on which our analysis focuses. 
$P(N_{ch})$ varies mildly up to $N_{ch}\sim 3500$, then decreases steeply.
By fitting it as a superposition of Gaussians, one can precisely reconstruct $\overline{N_{ch}}(b)$ and ${\rm Var}(N_{ch}|b=0)$~\cite{Yousefnia:2021cup} (Appendix~\ref{s:bayesian}). 
This fit is shown in Fig.~\ref{fig:panelplot} (a). 
The ``knee'' of the distribution, defined as the mean value of $N_{ch}$ for collisions at $b=0$, is reconstructed precisely, and indicated as a vertical line.   
The steep fall of $P(N_{ch})$ above the knee gives direct access to ${\rm Var}(N_{ch}|b=0)$. 
[Note that the variance is only reconstructed at $b=0$, and one must resort to assumptions as to its dependence on $b$. 
We have checked that our results are robust with respect to these assumptions, see Appendix~\ref{s:bayesian}  and  \ref{s:fits}.]
We refer to events above the knee as ultracentral collisions~\cite{Luzum:2012wu,CMS:2013bza}. 
They are a small fraction of the total number of events,  $0.35\%$, but ATLAS has recorded enough collisions that a few events are seen with values of $N_{ch}$ larger than the knee by $20\%$, corresponding to 4 standard deviations. 
Note that Poisson fluctuations contribute only by 15\% to the variance~\cite{Yousefnia:2021cup}, so that the fluctuations of $N_{ch}$ are mostly dynamical. 

We then model the fluctuations of $[p_t]$. 
In the same way as we have assumed that the probability of $N_{ch}$ at fixed $b$ is Gaussian, we assume that the joint probability of $N_{ch}$  and $[p_t]$, such as displayed in Fig.~\ref{fig:hydro}, is a two-dimensional Gaussian (Appendix~\ref{s:gaussian}).  
It is characterized by five quantities: 
The mean and variance of $[p_t]$ and $N_{ch}$, which we denote by $\overline{p_t}(b)$, $\overline{N_{ch}}(b)$, ${\rm Var}(p_t|b)$,  ${\rm Var}(N_{ch}|b)$, and the covariance or, equivalently, the Pearson correlation coefficient $r_{N_{ch}}(b)$ between $[p_t]$ and $N_{ch}$, which we expect to be positive as illustrated in Fig.~\ref{fig:hydro}. 
 $\overline{N_{ch}}(b)$ and ${\rm Var}(N_{ch}|b)$ are obtained from the fit to $P(N_{ch})$, as explained above. 
The mean transverse momentum is essentially independent of centrality for the 30\% most central collisions~\cite{ALICE:2018hza}, therefore, we assume that $\overline{p_t}(b)$ is independent of $b$, and we denote its value by $\overline{p_{t0}}$.
Since we only evaluate the fluctuations around $\overline{p_{t0}}$, results are independent of its value. 
The variance ${\rm Var}(p_t|b)$ may have a non-trivial dependence on the impact parameter, but a smooth one. 
For statistical fluctuations, it is proportional to $1/N_{ch}$.   
We allow for a more general power-law dependence ${\rm Var}(p_t|b)=\sigma_{\delta p_t}^2(\overline{N_{ch}}(0)/\overline{N_{ch}}(b))^{\alpha}$,
where $\sigma_{\delta p_t}$ and $\alpha$ are constants. 
Finally, we ignore the impact parameter dependence of the correlation coefficient $r_{N_{ch}}$ for simplicity. 

With this Gaussian ansatz, one can evaluate analytically the quantities entering the right-hand side of Eq.~(\ref{decomposition}) as a function of the parameters of the Gaussian (Appendix~\ref{s:gaussian}), and the averages over $b$ are evaluated using the probability distribution $P(b|N_{ch})$ obtained using the Bayesian method outlined above (Appendix~\ref{s:bayesian}). 
The remaining three parameters ($\sigma_{\delta p_t}$, $\alpha$ and $r_{N_{ch}}$) are fitted to ATLAS data. 

Let us first examine the distribution of $[p_t]$ and $N_{ch}$ returned by our fit, which is represented in the left panel of Fig.~\ref{fig:panelplot} (b). 
The white curves represent 99\% confidence ellipses at fixed impact parameter~\cite{Yousefnia:2021cup}. 
One sees that they are tilted with respect to the horizontal axis, as in the hydrodynamic calculation of Fig.~\ref{fig:hydro}. 
This tilt reflects the positive correlation between  $[p_t]$ and $N_{ch}$, parameterized by $r_{N_{ch}}$. 
As explained above, this correlation is a natural consequence of  thermalization. 
The width of the $[p_t]$ distribution for fixed $N_{ch}$ is due in part to the fact that several ellipses contribute for a given $N_{ch}$ (first term in Eq.~(\ref{decomposition})), and in part to  the vertical width of a single ellipse (second term in Eq.~(\ref{decomposition})). 

The left panel of Fig.~\ref{fig:panelplot} (c) displays the data and the model fit, as well as the two terms of Eq.~(\ref{decomposition}). 
The model explains precisely the observed decrease of the variance around the knee. 
It comes from the first term, namely, from impact parameter fluctuations at fixed $N_{ch}$, whose effect becomes negligible in ultracentral collisions. 
The magnitude of this term is essentially determined by the correlation coefficient $r_{N_{ch}}$, which is thus constrained by data. 

 As a corollary, we predict a small increase in the average transverse momentum, represented as a black line in Fig.~\ref{fig:panelplot} (b), in ultracentral collisions. 
This effect, which had been predicted a while ago~\cite{Gardim:2019brr,Nijs:2021clz}, has recently been observed by CMS collaboration~\cite{Bernardes:2023nnf}. 
Note that the increase is quantitatively predicted by our model calculation. 

A specificity of the ATLAS analysis is that it uses, in addition to $N_{ch}$, an alternative centrality estimator, which is the transverse energy $E_T$ (defined as energy multiplied by $\sin\theta$) deposited in two calorimeters located symmetrically on both sides of the collision point, which cover roughly the ranges $1^\circ<\theta< 5^\circ$ and $175^\circ<\theta< 179^\circ$. 
The analysis of the variance is repeated by sorting events according to $E_T$, rather than $N_{ch}$, and shown in the right panel of Fig.~\ref{fig:panelplot}.
In the same way, our model calculation can be repeated, replacing $N_{ch}$ with $E_T$ everywhere. 
This is a useful and non-trivial check of the validity of our approach. 
Even though the distributions of  $N_{ch}$ and $E_T$ look similar in shape (Fig.~\ref{fig:panelplot} (a)), the fall above the knee is steeper for $E_T$ than for $N_{ch}$, and there are only $0.26\%$ of events above the knee for $E_T$, as opposed to $0.35\%$ for $N_{ch}$.
It is interesting to notice that despite this significant difference, the decrease of the variance observed by ATLAS (Fig.~\ref{fig:panelplot} (c)) still occurs around the knee. 
The parameters $\sigma_{\delta p_t}$ and $\alpha$, which determine the dependence of the variance of $[p_t]$ on impact parameter, should not depend on whether one classifies events according to $N_{ch}$ or $E_T$. 
We determine the values that give the best simultaneous agreement with  $N_{ch}$ and $E_T$-based data (Appendix~\ref{s:fits}).  
The Pearson correlation coefficient $r_{E_T}$ between $[p_t]$ and $E_T$ need not coincide with $r_{N_{ch}}$ and is fitted independently. 
Note that $r_{N_{ch}}$ corresponds to the correlation between $[p_t]$ and $N_{ch}$ for the {\it same\/} particles, while $r_{E_T}$ represents the correlation between $[p_t]$ and the $E_T$ measured in  different angular windows. 
One therefore expects $r_{E_T}<r_{N_{ch}}$, which is confirmed by our fit. 
Values, however, are very similar, which shows that particle depositions in different $\theta$ windows are very strongly correlated. 

We have revealed a new effect of thermalization which involves the momenta of particles, rather than their direction. 
It is spectacular because of its unique centrality dependence
(as opposed to the presence of excess photons, also interpreted as resulting from thermal production~\cite{Wilde:2012wc,PHENIX:2014nkk}). 
Our study thus highlights the importance of impact parameter, which defines the geometry and is an essential ingredient of the hydrodynamic description:
The momentum per particle depends on its magnitude, much in the same way as elliptic flow is driven by its orientation. 
It is interesting to note that the impact parameter is a classical quantity, in the sense that its quantum uncertainty is negligible:
Heisenberg's principle gives $\delta b\equiv \hbar/P\sim 4\times 10^{-7}$~fm for a Pb+Pb collision at the LHC, negligible compared to the range spanned by $b$, of order $15$~fm.\footnote{Note that in event-by-event simulations, the impact parameter is correctly defined only if each nucleus is recentered after randomly drawing nucleon positions. 
The recentering correction is larger by orders of magnitude than the quantum uncertainty. 
It is {\it not\/} implemented in the simulations shown in Fig.~\ref{fig:hydro}, but this does not alter the conclusions drawn from this figure.}  
It is actually the {\it only\/} classical quantity characterizing a collision, and collisions with the same impact parameter differ only by quantum fluctuations. 
Due to the high energy of the collision, however, a single quantum fluctuation can produce a large number of particles, which promotes it to the status of a  classical fluctuation. 
(Elliptic flow in central collisions~\cite{PHOBOS:2006dbo} and triangular flow~\cite{Alver:2010gr} are driven by a similar mechanism.)
The effect studied in this paper involves a subtle interplay between classical fluctuations of impact parameter, and quantum fluctuations of the collision multiplicity. 

\section*{Acknowledgments}
We thank the Institute for Nuclear Theory at the University of Washington for hosting the program "Intersection of nuclear structure and high‐energy nuclear collisions" during which this work was initiated. 
We thank Govert Nijs and Wilke van der Schee for the discussions, and Jean-Paul Blaizot for useful comments on the manuscript.
R. S. is supported by the Polish National Science Center under grant NAWA PRELUDIUM BIS: PPN/STA/2021/1/00040/U/00001 and PRELUDIUM BIS: 2019/35/O/ST2/00357.
S.B and J.J are supported by DOE DE-FG02-87ER40331 and DE-SC0024602. M. L. thanks the S\~ao Paulo Research Foundation (FAPESP) for support under grants 2021/08465-9, 2018/24720-6, and 2017/05685-2, as well as the support of the Brazilian National Council for Scientific and Technological Development (CNPq).
We acknowledge support from the ``Emilie du Ch\^atelet'' visitor programme and from the GLUODYNAMICS project funded by the ``P2IO LabEx (ANR-10-LABX-0038)'' in the framework ``Investissements d'Avenir'' (ANR-11-IDEX-0003-01) managed by the Agence Nationale de la Recherche (ANR).

\appendix
\section{Bayesian reconstruction of impact parameter}
\label{s:bayesian}

We denote generically by $N$ the observable used as a centrality estimator, which can be either $N_{ch}$ or $E_T$. 
We assume that the distribution of $N$ at fixed $b$ is Gaussian: 
\begin{equation}
\label{probn}
P(N|b)=\frac{1}{\sqrt{2\pi{\rm Var}(N|b)}}\exp\left(-\frac{\left(N-\overline{N}(b)\right)^2}{2{\rm Var}(N|b)}\right).
\end{equation}
We introduce as an auxiliary variable the cumulative distribution of $b$~\cite{Das:2017ned}:
\begin{equation}
\label{defcb}
c_b=\int_0^b P(b')db'\simeq \frac{\pi b^2}{\sigma},
\end{equation}
where $P(b)\simeq 2\pi b/\sigma$ is the probability distribution of $b$, and $\sigma$ is the cross section of the nucleus-nucleus collision. 
$c_b$, which lies between $0$ and $1$, is usually called the centrality fraction.  
With this auxiliary variable, the probability distribution of $N$ can be written as $P(N)=\int_0^1 P(N|b)dc_b$. 
We assume that $\overline{N}(b)$ is a smooth function of $c_b$, which we parametrize as the exponential of a polynomial. 
A polynomial of degree $3$ is enough to obtain excellent fits to $P(N)$ in the chosen range: 
\begin{equation}
\overline{N}(b)=\overline{N}(0)\exp\left(-\sum_{i=1}^3 a_i (c_b)^i\right).
\end{equation}
Similarly, the variance  ${\rm Var}(N|b)$ is assumed to vary smoothly with $c_b$. 
By default, we assume that  ${\rm Var}(N|b)/\overline{N}(b)$ is constant. 
The parameters are fitted to the distribution $P(N_{ch})$ and $P(E_T)$  measured by ATLAS in Pb+Pb collisions. 
We normalize these probability distributions using  the centrality calibration provided by the ATLAS collaboration, that 40\% of events have $N_{ch}>705$. 
The fit is in agreement with data within 2\%. 
We have also tested two alternative scenarios, assuming either that ${\rm Var}(N|b)$ is constant or that the ratio  ${\rm Var}(N|b)/\overline{N}(b)^2$ is constant. 
The quality of the fit is as good and the fit parameters are essentially unchanged, as shown in Table~\ref{fitparameters}.  
\begin{table}
\begin{tabular}{|c||c|c|}
\hline
$N$&$N_{ch}$&$E_T$\cr
\hline
$\overline{N}(b=0)$&$3683\pm 4$&$4.435\pm 0.003$~TeV\cr
$\sqrt{{\rm Var}(N|b=0)}$&$168.1\pm 0.1$&$0.1433\pm 0.0001$~TeV\cr
$a_1$&$4.31\pm 0.02$&$4.18\pm 0.01$\cr
$a_2$&$-4.19\pm 0.03$&$-3.45\pm 0.01$\cr
$a_3$&$10.21\pm 0.09$&$8.54\pm 0.05$\cr
\hline
\end{tabular}
\caption{\label{fitparameters} 
Values of fit parameters for Pb+Pb collisions at center-of-mass energy $5.02$~TeV per nucleon pair. 
For each parameter, the central value is that obtained by assuming that the variance is proportional to the mean, and the error bars reflect the changes when one assumes instead that the variance is constant, or proportional to the square of the mean. 
}
\end{table}
The largest source of error in extracting impact parameter from data is a global normalization, since it is difficult to evaluate experimentally which fraction of the cross-section is seen in detectors~\cite{ALICE:2013hur}. 
We ignore this issue here, since we are interested in ultracentral collisions. 
When we write that we use the 20\% most central events, we mean that we use the 20\% most central of the events that are actually seen in the detector. 
The overlapping circles in Fig.~2 (b) are a schematic representation of the colliding Pb nuclei, with radius $R=6.62$~fm. 
The values of $b$ are calculated assuming that the inelastic cross section of Pb+Pb collisions is $767$~fm$^2$. 

The averages over impact parameter in Eq.~(1) of the paper are evaluated by using $c_b$ as an integration variable, rather than $b$. 
Its probability distribution at fixed $N$ is given by Bayes' theorem:
\begin{equation}
P(c_b|N)=\frac{1}{P(N)}P(N|b),
\end{equation}
where we have used the fact that the probability distribution of $c_b$ is uniform, $P(c_b)=1$.
$P(c_b|N)$ becomes narrower in ultracentral collisions, as illustrated in Ref.~\cite{Das:2017ned}.

\section{Simulations with hydrodynamics and HIJING}
\label{s:hydro}

The setup of our hydrodynamic calculation is identical to that of Ref.~\cite{Bozek:2021mov}. 
We use a boost-invariant version of the hydrodynamic code MUSIC~\cite{Paquet:2015lta} with the default freeze-out temperature $T_f=135$~MeV. 
We assume a constant shear viscosity to entropy density ratio $\eta/s=0.12$, and the bulk viscosity is set to zero.
The initial entropy distributions are taken from the TRENTO model~\cite{Moreland:2014oya}, where the parameters are fixed as follows.
The most important parameter is the parameter $p$ which defines the dependence of the density on the thickness functions of incoming nuclei, which is set to $p=0$, corresponding to a geometric mean, which is the default choice.
The parameter defining the strength of multiplicity fluctuation is set to $k=2.5$ (the default being $k=1$).
With this choice, the relative multiplicity fluctuation is compatible (within statistical errors) with ATLAS data in Table~\ref{fitparameters}.
The nucleon-nucleon cross section is set to $\sigma_{NN}=7.0$~fm$^2$ (instead of the default $\sigma_{NN}=6.4$~fm$^2$).

The normalization of the entropy density from the TRENTO model is adjusted so as to reproduce the charged multiplicity measured by ALICE in Pb+Pb collisions at $5.02$~TeV~\cite{ALICE:2015juo}. 
Despite this normalization, the average mulplicity is $\overline{N_{ch}}=6660\pm 30$, much larger than that seen by ATLAS (Table~\ref{fitparameters}). 
The main reason is that some of the particles escape detection, even within the specified angular and $p_t$ range, and the data are not corrected for the reconstruction efficiency. 
In addition, we expect deviations between the model and data for two reasons. 
First, hydrodynamic models typically underestimate the pion yield at low $p_t$~\cite{Grossi:2021gqi,Guillen:2020nul}. 
Since the calculation is adjusted to reproduce the total charged multiplicity, which is dominated by pions, this implies in turn that it should overestimate the yield for $p_t>0.5$~GeV/c, which is the range where it is measured by ATLAS. 
Second, our hydrodynamic calculation assumes that the momentum distribution is independent of rapidity. 
In reality, it is maximum near mid-rapidity, in the region covered by the ALICE acceptance. 
This should also lead to slightly overestimating the multiplicity seen by ATLAS, whose inner detector covers a broader range in rapidity. 

The width of $p_t$ fluctuations from our hydrodynamic calculation is $\sigma_{\delta p_t}=13\pm 1$~MeV$/c$.
Note that they are dynamical fluctuations only. 
The reason is that we do not sample particles according to a Monte Carlo algorithm, but simply calculate the expectation value of $[p_t]$ at freeze-out.
Therefore, the width of $[p_t]$ fluctuations from the hydrodynamic calculation can in principle be compared directly with that measured experimentally. 
Our value is somewhat higher than the value $\sigma_{\delta p_t}=9.357$~MeV$/c$ inferred from ATLAS data (see Fig.~2 (c) of the paper). 
The fact that hydrodynamics overestimates $[p_t]$ fluctuations is an old problem~\cite{Bozek:2012fw}, which can be remedied by carefully tuning the fluctuations of the initial density profile~\cite{Bozek:2017elk,Bernhard:2019bmu,Everett:2020xug}.  
It is the reason why we choose to fit the magnitude of $[p_t]$ fluctuations to data, rather than obtain it from a hydrodynamic calculation. 

The Pearson correlation coefficient between $N_{ch}$ and $[p_t]$ from our hydrodynamic calculation is $r_{N_{ch}}=0.61\pm 0.08$ compatible with the value $r_{N_{ch}}=0.676$ returned by the fit to ATLAS data (Fig.~2 (c) of the paper). 

Simulations with HIJING shown in Fig.~3 of the paper follow the same setup as in Ref.~\cite{Bhatta:2021qfk}. 
The average multiplicity is $\overline{N_{ch}}=5149$, somewhat lower than in the hydrodynamic calculation, and the average value of $p_t$, denoted by $\overline{p_t}$, is $941$~MeV$/c$, also lower than in the hydrodynamic simulation ($\overline{p_t}=1070$~MeV$/c$). 

\section{Distribution of $[p_t]$ and $N$}
\label{s:gaussian}

We assume that the probability distribution of $[p_t]$ and the centrality estimator $N$ at fixed $b$ is a two-dimensional Gaussian. 
\begin{eqnarray}
\label{probptn}
P(\delta p_t,N)&=&\frac{1}{2\pi\sqrt{(1-r^2){\rm Var}(p_t){\rm Var}(N)}}\cr
&&\times\exp\left(\frac{1}{1-r^2}\left(-\frac{(\delta p_t)^2}{2{\rm Var}(p_t)}
-\frac{\left(N-\overline{N}\right)^2}{2{\rm Var}(N)}
\right.\right.\cr&&\left.\left.
+\frac{r\left(N-\overline{N}\right)\delta p_t}{\sqrt{{\rm Var}(N){\rm Var}(p_t)}}
\right)\right),
\end{eqnarray}
where we have omitted the dependence on $b$ to simplify the expression, and introduced the shorthand $\delta p_t\equiv [p_t]-\overline{p_t(b)}$. 

The linear correlation between $[p_t]$ and $N$ is
\begin{equation}
\label{linearcorr}
\int{\delta p_t(N-\overline{N})P(\delta p_t,N)dNd\delta p_t}=r\sqrt{{\rm Var}(N){\rm Var}(p_t)},
\end{equation}
where integrations on both variables are from $-\infty$ to $+\infty$.

A property of the two-dimensional Gaussian distribution is that its marginal distributions, obtained upon integrating over one of the variables, are also Gaussian. 
Integrating (\ref{probptn}) over $\delta p_t$, one recovers Eq.~(\ref{probn}). 
Integrating (\ref{probptn}) over $N$, one obtains similarly: 
\begin{equation}
\label{probpt}
P(\delta p_t|b)=\frac{1}{\sqrt{2\pi{\rm Var}(p_t|b)}}\exp\left(-\frac{(\delta p_t)^2}{2{\rm Var}(p_t|b)}\right),
\end{equation}
where we have restored the dependence on $b$. 

Another property of the two-dimensional Gaussian distribution is that if one fixes one of the variables, e.g. $N$, the probability of the other variable, e.g. $\delta p_t$, is also Gaussian. 
Its centre is: 
 \begin{equation}
  \label{mean}
       \overline{\delta p_t}(N,b)=r(b)\sqrt{\frac{{\rm Var}(p_t|b)}{{\rm Var}(N|b)}}\left(N-\overline{N}(b)\right).
\end{equation}
It increases linearly with $N$ due to the positive correlation, as exemplified in Fig.~3 of the paper. 
On the other hand, the variance of the distribution of $\delta p_t$ at fixed $N$ is independent of $N$: 
\begin{equation}
  \label{var}
        {\rm Var}(p_t|N,b)= \left(1-r(b)^2\right) {\rm Var}(p_t|b).
\end{equation}
This equation expresses that by fixing the value of $N$, one narrows the distribution of $\delta p_t$ due to its positive correlation with $N$.

\section{Fitting the variance of $[p_t]$ fluctuations}
\label{s:fits}

ATLAS provides us with two data sets for the centrality dependence of the variance, depending on whether centrality is determined with $N_{ch}$ or $E_T$.
We first carry out a standard $\chi^2$ fit for each of these sets, where the error is the quadratic sum of the statistical and systematic errors on the data points. 
The three fit parameters are $\sigma_{\delta p_t}$ (the standard deviation of $[p_t]$ for $b=0$), $\alpha$ (which defines the decrease of the variance as a function of impact parameter), and the Pearson correlation coefficient $r$ between $[p_t]$ and the centrality estimator for fixed $b$. 
Consistency of our model requires that $\sigma_{\delta p_t}$ and $\alpha$, whose definition does not involve the centrality estimator, are identical for $N_{ch}$ and $E_T$ based data for a given $p_t$ selection. 
Values of $\sigma_{\delta p_t}$ are identical within less than $1\%$, but values of $\alpha$ differ by $6\%$, with $E_T$-based data favoring a larger $\alpha$. 
We then fix the values of $\sigma_{\delta p_t}$ and $\alpha$ to the average values of $N_{ch}$ and $E_T$-based results, and redo the fits by fitting solely the Pearson correlation coefficient $r$ for each of the two data sets. 
Due to the small tension between the values of $\alpha$,  our fit slightly overestimates the variance for the lowest values of $N_{ch}$, and slightly underestimates it for the lowest values of $E_T$. 
This effect is of little relevance to our study which focuses on ultracentral collisions, and we have not investigated its origin. 

The values of $\alpha$ are close to $1.2$, which implies that the variation of dynamical fluctuations with impact parameter is faster than that of statistical fluctuations, for which $\alpha=1$.  
$\sigma_{\delta p_t}$ is close to $10$~MeV$/c$, while the average value of $p_t$ is close to $1$~GeV$/c$. 
This corresponds to a relative dynamical fluctuation of order $1\%$ in central collisions. 
The values of the Pearson correlation coefficient end up being similar, between $0.6$ and $0.7$, for both data sets. 

The results shown are obtained by assuming that the variance of the charged multiplicity is proportional to the mean, that is,  ${\rm Var}(N|b)/\overline{N}(b)$ is constant. 
As explained in Sec.~\ref{s:bayesian}, we have also tested two alternative scenarios, assuming either that ${\rm Var}(N|b)$ is constant or that the ratio  ${\rm Var}(N|b)/\overline{N}(b)^2$ is constant. 
We have checked that the fit to the data is as good.  
The values of fit parameters vary only by 3\% for $\alpha$, and even less for $\sigma_{\delta p_t}$ and  $r$.

\end{document}